\documentclass[aps,prl,reprint,superscriptaddress,nofootinbib,floatfix]{revtex4-2}
\usepackage{amsmath,amssymb,graphicx,bm,xcolor}
\usepackage{needspace}
\usepackage[colorlinks=true,linkcolor=blue,citecolor=blue,urlcolor=blue]{hyperref}
\newcommand{\dd}{\mathrm{d}}
\newcommand{\ii}{\mathrm{i}}
\newcommand{\Ec}{E_{\mathrm c}}
\newcommand{\wrot}{\omega_{\mathrm{ro}}}

\begin{document}
\title{Quasi-normal Modes and Instabilities of Rotating Relativistic Stars}
\author{Yong Gao}
\email{yong.gao@aei.mpg.de}
\affiliation{Max Planck Institute for Gravitational Physics (Albert Einstein
Institute), 14476 Potsdam, Germany}
\author{Masaru Shibata}
\affiliation{Max Planck Institute for Gravitational Physics (Albert Einstein
Institute), 14476 Potsdam, Germany}
\affiliation{Center of Gravitational Physics and Quantum Information, Yukawa Institute
for Theoretical Physics, Kyoto University, Kyoto 606-8502, Japan}
\date{October 24, 2026}

\begin{abstract}
We derive the basic equations suitable for the coupled fluid--spacetime eigenvalue problem from the
second-order action in full general relativity and accurately determine quasi-normal
modes of a variety of rapidly rotating stars with outgoing-wave boundary conditions.
For MIT-bag strange stars, the spectrum realizes Chandrasekhar's
instability picture, with distinct Chandrasekhar--Friedman--Schutz and Jacobi-like
neutral conditions and the transition to dynamical bar-mode growth. For neutron
stars, we determine the $f$-, $r$-, and other inertial-mode spectra and their radiative
damping or growth times. Interactions between $f$- and inertial modes reveal
avoided crossings and exchanges for radiative damping or growth rates.
We locate quasiradial collapse thresholds through a static
stability analysis and independently verify them by tracking the fundamental
$F$-mode frequency to zero, clarifying that the central density at the onset of the dynamical instability is always larger than that of the secular instability for uniform rotation.
Our formalism and numerical implementation can also resolve highly damped $w$-modes.
\end{abstract}
\maketitle

\textit{Introduction.---}
Quasi-normal modes describe the oscillations and gravitational-wave (GW)
emission of rotating stars and determine their dynamical and secular
stabilities~\cite{Kokkotas:1999bd,Friedman:2013xza}.
The stability of rotating self-gravitating bodies has long been studied
through their oscillation spectra, as in Chandrasekhar's theory of
ellipsoidal figures \cite{Chandrasekhar:1969kvx}.
General relativity (GR) adds gravitational radiation, which can damp an
oscillation or drive secular growth through the
Chandrasekhar--Friedman--Schutz (CFS) mechanism
\cite{Chandrasekhar:1970jaz,Friedman:1975gjz,Friedman:1978hf,Friedman:1978wla}.
The $r$-mode instability can influence the spin evolution of neutron stars (NSs)
and produce continuous GWs
\cite{Andersson:1997xt,Friedman:1997uh,Owen:1998xg}, while the $f$-mode
instability is of particular interest in hot, rapidly rotating newborn
NSs and long-lived remnants of binary NS mergers
\cite{Lai:1994ke,Gaertig:2011bm,Doneva:2015jaa}.
Rapid rotation can also drive a dynamical bar-mode instability without
dissipation \cite{Chandrasekhar:1969kvx,Shibata:2000jt}, while
differential rotation permits the corotation instability
\cite{Centrella:2001xp,Shibata:2002mr,Shibata:2003yj,Saijo:2002nmv,Watts:2003nn,Saijo:2005gb,Passamonti:2014eea},
which may also operate in merger remnants
\cite{Paschalidis:2015mla,Radice:2016gym,Gao:2025nfj}.

GWs from a binary NS inspiral were detected in GW170817
\cite{LIGOScientific:2017vwq}, and dedicated searches target $r$-mode
emission from rotating pulsars
\cite{Fesik:2020tvn,Rajbhandari:2021pgc,LIGOScientific:2021yby}.
Detecting stellar oscillations and instabilities would provide
complementary information about internal dynamics and dense matter
\cite{Andersson:1996pn,Andersson:1997rn,Glampedakis:2017nqy,ET:2019dnz}.
Predicting and interpreting these signals require both oscillation
frequencies and damping or growth rates, encoded in the real and
imaginary parts of the quasi-normal mode spectrum.

The complex spectrum of spherical relativistic stars was established
through radial boundary-value calculations
\cite{Thorne:1967chi,Price:1969ufa,Thorne:1969khk} and subsequent
variational and numerical formulations
\cite{Detweiler:1973zza,Lindblom:1983ps,Detweiler:1985zz}.
Fluid-mode frequencies and instability boundaries have since been mapped
in Newtonian gravity \cite{Ipser:1990iwb,Ipser:1991qcv}, slow-rotation
expansions \cite{Kojima:1992ie,Kojima:1993zuv,Lockitch:2002sy}, the
Cowling approximation \cite{Yoshida:1997wn,Gaertig:2008uz,Doneva:2013zqa},
and calculations assuming conformal flatness \cite{Servignat:2024loh,Jaraba:2026nmq}.
In full GR, static calculations locate secular neutral
points \cite{Stergioulas:1997ja,Morsink:1998db}, while time evolutions
determine frequencies and $f$-mode CFS thresholds
\cite{Font:2001ew,Zink:2010bq,Kruger:2019zuz} and follow dynamical bar
formation \cite{Shibata:2000jt,Baiotti:2006wn}.
Rapid rotation couples angular multipoles and requires a
two-dimensional solution with outgoing radiation~\cite{Priou:1992zxa,Kruger:2020ykw}.
Radiative damping and growth rates at rapid rotation have largely been
estimated from mode energies and multipole emission formulae
\cite{Gaertig:2010kc}.

Here we present, to our knowledge, the first direct calculation of the
complex frequencies of $f$, $r$, other inertial, and strongly damped $w$
modes of rapidly rotating relativistic stars in full GR.
We obtain these modes by applying a spectral Ritz--Galerkin method to the
second-order Einstein--fluid action in Lorenz gauge on highly accurate
numerical backgrounds, with exterior complex scaling enforcing outgoing
radiation. The detailed formalism will be presented in a separate
paper~\cite{Gao:2026prep}.
We use geometrized units $G=c=1$. $g_{ab}$ denotes the spacetime metric and $\nabla_a$ denotes the covariant derivative associated with $g_{ab}$.

\textit{Formulation.---}
We consider a linear perturbation of a stationary, axisymmetric
star composed of a zero-temperature, uniformly rotating perfect fluid with four-velocity
$u^a=u^t(t^a+\Omega\phi^a)$, where $t^a$ and $\phi^a$ are the stationary
and axial Killing vectors, $\Omega$ is the angular velocity measured at
infinity, and $u^t$ is fixed by $u^au_a=-1$.
The equation of state (EOS) specifies the total energy density
$\epsilon(n)$ as a function of the proper baryon number density $n$, with pressure
$p=n\,\dd\epsilon/\dd n-\epsilon$.
We compute the background equilibria to high accuracy with a code
developed using the KADATH spectral library~\cite{Grandclement:2009ju}.

We describe a perturbation by the Eulerian metric change
$h_{ab}=\delta g_{ab}$ and the Lagrangian displacement $\xi^a$ of the
fluid. The Eulerian perturbation $\delta$ compares fields at the same
spacetime point, whereas the Lagrangian perturbation $\Delta$ follows
the same fluid element. They are related by
$\Delta=\delta+\mathcal L_\xi$, where $\mathcal L_\xi$ is the Lie derivative
along $\xi^a$ \cite{Friedman:1978plf,Friedman:1978wla,Friedman:2013xza}.
Baryon conservation relates the density change to the Lagrangian
metric change~\cite{Friedman:1978wla}
\begin{equation}
 \Delta g_{ab}=h_{ab}+2\nabla_{(a}\xi_{b)}\,,\qquad
 \frac{\Delta n}{n}=-\frac12(g^{ab}+u^au^b)\Delta g_{ab}\,,
 \label{eq:material}
\end{equation}
so the density perturbation is determined by $h_{ab}$ and $\xi^a$.
We choose the displacement representative with $\xi^t=0$, leaving
three fluid components $\xi^i$.

For each mode, the perturbation amplitudes depend on $(r,\theta)$ and
have the harmonic dependence
\begin{equation}
 (h_{ab},\xi^i)\propto e^{\ii(\sigma t+m\phi)}\,,
 \qquad \wrot=\omega_{\mathrm{in}}+m\Omega\,.
 \label{eq:convention}
\end{equation}
Here $m$ is the integer azimuthal index,
$\sigma=\omega_{\mathrm{in}}+\ii\sigma_I$ is the complex angular frequency in the
inertial frame, $\omega_{\mathrm{in}}=\operatorname{Re}\sigma$ is its real part,
and $\wrot$ is the real frequency in the corotating frame.
The amplitude varies as $e^{-\sigma_I t}$, so $\sigma_I>0$ denotes
damping and $\sigma_I<0$ growth, with e-folding time
$\tau=|\sigma_I|^{-1}$.

We choose the Lorenz gauge $Z_b=\nabla^a\tilde h_{ab}=0$, where
$\tilde h_{ab}=h_{ab}-g_{ab}h^c{}_c/2$.
In this gauge, all ten linear metric equations have the wave operator
$\nabla^c\nabla_c\tilde h_{ab}$ as their principal part, with curvature
couplings and fluid source terms, as described in
Refs.~\cite[Sec.~7.1]{Friedman:2013xza} and
\cite[Sec.~II~A]{Kruger:2020ykw}.

Following Friedman and Schutz, we obtain the coupled perturbation
equations by varying the second-order Einstein--fluid action
\cite{Friedman:1975yjh,Friedman:1978wla,Friedman:2013xza}.
We include the displaced stellar surface and add a quadratic
gauge-fixing term proportional to $Z_aZ^a$. Variation with respect to
the metric and displacement gives the ten Lorenz-reduced linearized Einstein
equations and three fluid equations. After integration by parts, the
action contains at most first derivatives of the perturbations.

We then apply a spectral Ritz--Galerkin method
\cite{Ritz:1909rwo,Gander:2012kuf,Canuto:2006uzp,Meltzer1966ApJ,Bardeen1966ApJ}, expanding the ten components
of $\tilde h_{ab}$ and three components of $\xi^i$ in spectral bases and
collecting their coefficients into a vector $\mathbf x$.
Stationarity of the discretized action gives the matrix
eigenvalue problem
\begin{equation}
\left(\mathbf K_0+\sigma\mathbf K_1+\sigma^2\mathbf K_2\right)\mathbf x=0\,.
 \label{eq:pencil}
\end{equation}
The coefficient matrices $\mathbf K_0$, $\mathbf K_1$, and $\mathbf K_2$
come from terms with zero, one, and two time derivatives, respectively.

We impose regularity at the center and along the rotation axis, match the interior
and exterior metric perturbations at the stellar surface, and require
$\Delta p=0$ there. The exterior solution must contain only outgoing
radiation \cite{Andersson:1995wu,Gao:2025aqo}.
We impose this condition by exterior complex scaling \cite{Simon:1979hdk,Andersson:1995wu}.
Beyond a radius $r_b$ outside the star, we deform the radial integration path
used to assemble the exterior matrices to $r=r_b+s e^{\ii\vartheta}$,
where $s\geq0$ is real and $\vartheta$ is fixed.
The integration variables are $s$ and the real polar angle $\theta$,
with $\dd r=e^{\ii\vartheta}\dd s$.
We choose $\vartheta$ such that $\operatorname{Im}(\sigma e^{\ii\vartheta})<0$,
so the outgoing wave $e^{-\ii\sigma r}$ decays along this path.

For the conservative problem, the same action defines the inertial-frame
canonical energy $\Ec$ and canonical angular momentum $J_{\mathrm c}$
\cite{Friedman:1978plf,Friedman:1978hf,Friedman:1978wla}.
Outgoing GWs carry positive energy and lower
$\Ec$, allowing a negative-energy mode to grow through the CFS mechanism.
Viscosity instead lowers the corotating-frame canonical energy
$E_{\mathrm{c},r}=\Ec-\Omega J_{\mathrm c}$ and can drive the Jacobi-like
instability when $E_{\mathrm{c},r}<0$. The corresponding $f$-mode neutral
patterns are stationary in the inertial or corotating frame,
$\omega_{\mathrm{in}}=0$ or $\wrot=0$, respectively
\cite{Chandrasekhar:1970jaz,Gondek-Rosinska:2002rqd,Friedman:2013xza}.

\begin{figure}[t]
 \centering
 \includegraphics[width=\columnwidth]{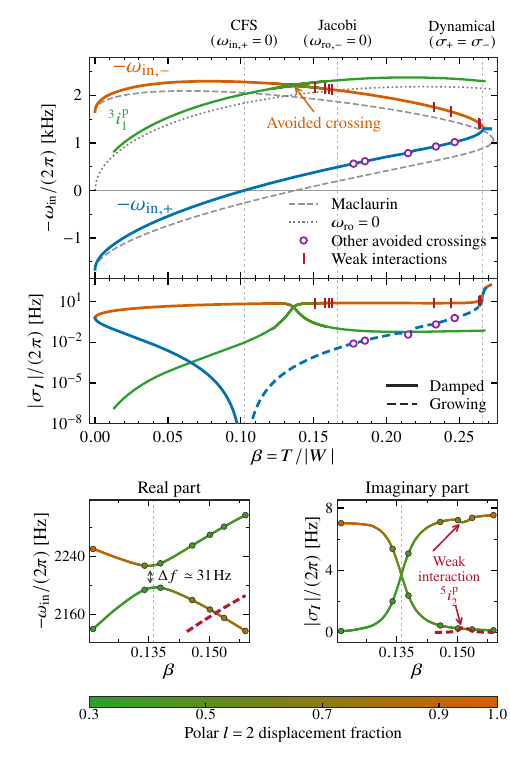}
 \vspace{-8mm}
 \caption{Quadrupolar mode frequencies of the MIT-bag sequence with $M_0=1.57\,M_\odot$ and $m=2$.
Top: signed real frequencies, $-\omega_{\mathrm{in}}/(2\pi)$, together with a Maclaurin comparison
normalized to the nonrotating $f$ frequency (dashed curves). Green shows the polar-led inertial
partner ${}^{3}i^{\mathrm p}_1$; purple circles mark the other avoided crossings
and short red ticks mark weak interactions.
Middle: radiation-rate magnitudes, with solid/dashed curves for damping/growth.
Bottom: real frequencies (left) and imaginary-part magnitudes
(right) at the first avoided crossing, $\beta\simeq0.1362$, with a minimum gap of
approximately 31 Hz. The pair's colors show the polar $l=2$ displacement
fraction; the red dashed curve shows the weak partner
${}^{5}i^{\mathrm p}_2$ near $\beta\simeq0.151$.
Vertical guides mark
inertial-frame neutrality, rotating-frame neutrality, and the extrapolated
conservative bar-mode collision.}
 \label{fig:qs}
\end{figure}

\textit{Instabilities of a self-bound star.---}
Chandrasekhar's analysis of Maclaurin spheroids provides the classical
picture of secular and dynamical bar-mode instabilities
\cite{Chandrasekhar:1969kvx} (see, e.g.,
Refs.~\cite{Lai:1993ve,Lai:1994hf,Andersson:2002ch}).
Among relativistic stars, self-bound strange stars can reach the large
rotational deformations needed to realize this picture
\cite{Gourgoulhon:1999vx,Gondek-Rosinska:2003zdo,Chen:2023bxx}.

We use the MIT-bag EOS $p=(\epsilon-4B)/3$ \cite{Farhi:1984qu}, with
$B=52.4\,\mathrm{MeV/fm^3}$, and follow a uniformly rotating
sequence at fixed baryon mass $M_0=1.57\,M_\odot$.
The nonrotating star has gravitational mass $M=1.387\,M_\odot$ and
radius $R=11.32$~km. Figure~\ref{fig:qs} follows the two $f$ branches
$\sigma_\pm$, labeled by the signs of $\omega_{\mathrm{in}}$ in the nonrotating
limit. The frequency panels show $-\omega_{\mathrm{in}}/(2\pi)$.

\begin{figure*}[t]
 \centering
 \includegraphics[width=0.95\textwidth]{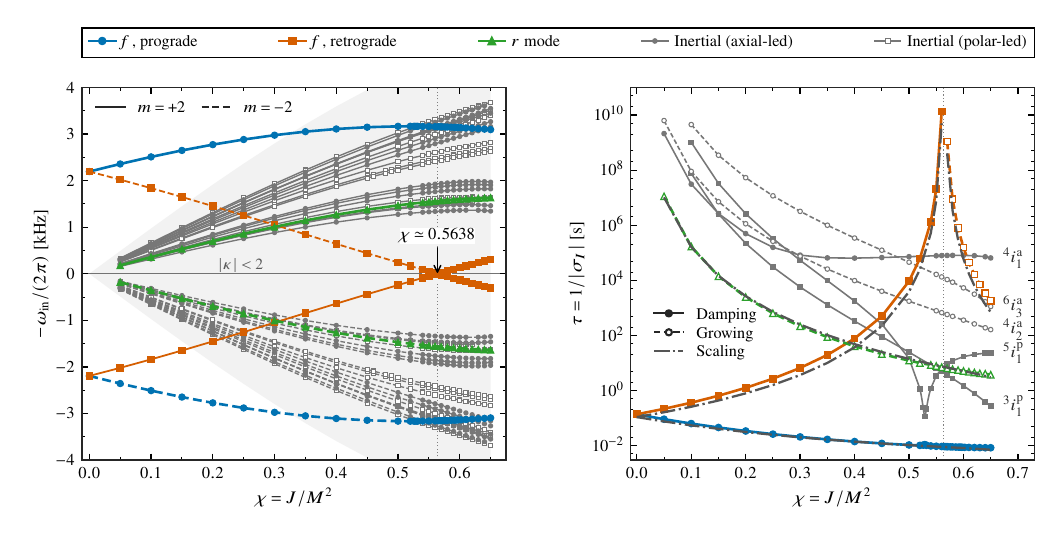}
 \vspace{-5mm}\caption{Fluid modes of the APR4 sequence with $M_0=2\,M_\odot$. Left: signed
inertial-frame frequency $-\omega_{\mathrm{in}}/(2\pi)$ against
$\chi=J/M^2$. Solid curves show $m=2$ and dashed curves the partners
obtained from $(\sigma,m)\mapsto(-\sigma^*,-m)$. The 18 additional
inertial branches are plotted in gray, with circles/open squares denoting
axial-led/polar-led modes.
Shading marks $|\kappa|<2$, with $\kappa=\wrot/\Omega$ in each $m$
sector. The $f$-mode neutral crossing is at $\chi\simeq0.5638$. {Right: damping
and growth times $\tau=|\sigma_I|^{-1}$ for the $f$ and $r$ modes and five selected inertial branches (gray)},
shown once for each conjugate pair, with filled/open markers and
solid/dashed curves for damping/growth. Dark gray dash-dotted curves give the approximate
$f/r$ radiation times described in the text. {For both panels, blue, orange, and green show the
prograde $f$, initially retrograde $f$, and $r$ branches, respectively.}}
 \label{fig:fluid}
\end{figure*}

Writing $\beta=T/|W|$ where $T$ and $W$ denote the rotational kinetic and gravitational potential energies, respectively, we find the CFS neutral point at $\beta=0.1023$,
where the initially retrograde branch becomes stationary in the inertial
frame and turns unstable. The Jacobi-like neutral point occurs at
$\beta=0.1665$, where the other branch becomes stationary in the rotating
frame. Relative to their common Maclaurin value, $\beta=0.1375$, the CFS
onset shifts downward \cite{Stergioulas:1997ja,Kruger:2019zuz},
whereas the Jacobi-like onset shifts upward \cite{Gondek-Rosinska:2002rqd}.
Viscosity is not included in the present study, so the latter branch remains radiatively damped.

At higher rotation, the two $f$ branches approach the dynamical bar-mode
instability. Its conservative onset is a Krein collision, in which two
real-frequency modes of opposite canonical energy coalesce and form a
growing and a decaying pair \cite{MacKay:1986hiq}.
Extrapolating the squared splitting of the outgoing roots places the
conservative collision near $\beta\simeq0.2661$, slightly below the
Maclaurin threshold $0.2738$. Radiation makes the splitting complex,
so this estimates the collision location without establishing an
exact degeneracy on the real sequence. By $\beta=0.2715$, the growing branch has
$\tau\simeq1.0$~ms. Such dynamical bar formation has also been followed in
fully relativistic simulations of differentially rotating NSs
\cite{Shibata:2000jt,Baiotti:2006wn,Loffler:2014jma,DePietri:2014mea}
and weakly differentially rotating strange stars~\cite{Zhou:2021upu}.

\Needspace{3\baselineskip}
\textit{Interactions with inertial modes.---}
Inertial modes are fluid oscillation modes restored by the Coriolis force
\cite{Lockitch:1998nq,Lockitch:2002sy}.
In our labels ${}^{\ell_0}i_n^{\mathrm{p/a}}$, $i$ denotes an inertial mode,
$\ell_0$ the highest spherical-harmonic degree of its uniform-density
Newtonian reference mode, and $n$ orders the reference roots by increasing
$\omega_{\mathrm{ro}}/\Omega$ within each family.
The superscripts p/a indicate whether the lowest nonzero harmonic is
polar or axial \cite{Lockitch:2000aa}.

We interpret their encounters with $f$ modes \cite{Barker:2016aka,Dewberry:2022olp}
using Heiss's two-mode model \cite{Heiss:1999alz}.
We write the frequencies before coupling as $\sigma_j=\omega_j+\ii\gamma_j$,
with $j=1$ for the $f$ mode and $j=2$ for the inertial mode.
$\omega_j$ and $\gamma_j$ are the real and imaginary parts in the
inertial frame, with $\gamma_j>0$ for damping and $\gamma_j<0$ for growth.
Near each encounter, we approximate the coupled frequencies by
\begin{equation}
 \lambda_\pm=\frac{\sigma_1+\sigma_2}{2}\pm
 \sqrt{\frac{(\sigma_1-\sigma_2)^2}{4}+V^2}\,,
 \label{eq:mode-interaction}
\end{equation}
where $V>0$ is the coupling strength.
At $\omega_1=\omega_2$, the splitting is $2\sqrt{V^2-g^2}$,
with $g=|\gamma_1-\gamma_2|/2$.
Within this model, strong interactions ($V>g$) produce the usual avoided crossing: the real
parts repel and the imaginary parts cross.
For weak interactions ($V<g$), the real parts cross and the
imaginary parts avoid crossing.
The eigenfunctions mix in both cases, exchanging their $f$-like and
inertial character only in the strong case.

Figure~\ref{fig:qs} shows a strong encounter with ${}^{3}i^{\mathrm p}_1$
near $\beta=0.1362$, with a real-frequency gap of $31$~Hz.
The weak encounter with ${}^{5}i^{\mathrm p}_2$ near $\beta=0.151$
has a smallest sampled imaginary-part gap of $6.8$~Hz.
The upper $f$-like branch dips while the lower inertial branch peaks in
the imaginary-part panel.

\begin{figure}[!tb]
 \centering
 \includegraphics[width=\columnwidth]{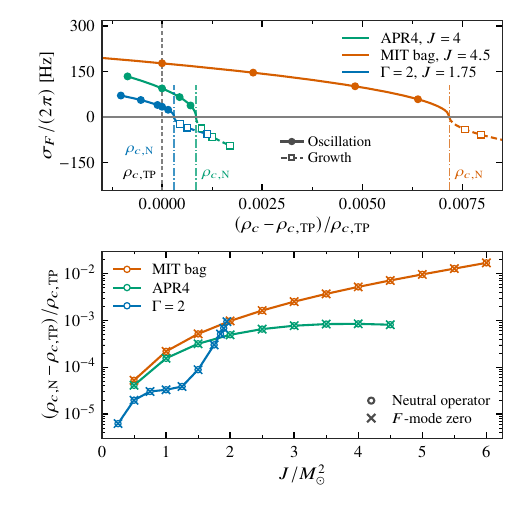}
 \vspace{-10mm}
 \caption{Axisymmetric $F$-mode neutrality. Top: {transition from
oscillation (solid, circles) to nonoscillatory growth (dashed, open squares)
at fixed $J$ (in units of $M_\odot^2$).}
Markers show computed modes; curves interpolate
$\operatorname{Re}(\sigma_F^2)$. Gray and colored vertical lines mark
the turning point (TP) and neutral-operator zeros, respectively.
Bottom: relative central-density offsets from the turning point,
determined by the neutral operator (circles) and the vanishing
$F$-mode frequency (crosses).}
 \label{fig:neutral}
\end{figure}

\textit{Neutron-star fluid modes.---}
For the neutron-star sequence, we use a smoothed representation of the four-segment
APR4 EOS \cite{Akmal:1998cf,Read:2008iy}
and fix the baryon mass to $M_0=2\,M_\odot$.
The nonrotating star has $M=1.743\,M_\odot$,
$R=11.11$~km, and complex $f$-mode frequencies
$\sigma/(2\pi)=(\pm2.196+0.001171\ii)$~kHz.
Figure~\ref{fig:fluid} follows the sequence for $ 0\leq \chi(=J/M^2) \leq 0.65$.

On the prograde $f$ branch, the oscillation frequency
$|\omega_{\mathrm{in}}|/(2\pi)$ rises to $3.169$~kHz near
$\chi=0.52$ and then decreases. The damping time falls overall
from $0.136$~s to $8.3$~ms, with a local increase at the inertial-mode
encounter discussed below. On the initially retrograde branch, the
real frequency $\omega_{\mathrm{in}}/(2\pi)$ decreases monotonically and crosses zero
at $\chi=0.5638$.
For $m=2$, setting $\sigma=0$ in Eq.~\eqref{eq:pencil} gives the static
equation $\mathbf K_0\mathbf x=0$.
This static calculation independently gives the same CFS threshold.

{The $r$ mode is a special inertial mode whose Newtonian $l=m=2$
branch has purely toroidal motion at leading order in rotation
\cite{Papaloizou:1978zz,Lockitch:1998nq}.
Its continuation to relativistic barotropic stars contains polar components
and is an axial-led hybrid
\cite{Lockitch:2000aa,Lockitch:2002sy}.
Its pattern is retrograde in the corotating frame but prograde in the
inertial frame, making it CFS unstable down to arbitrarily slow rotation
in the absence of viscosity \cite{Andersson:1997xt,Friedman:1997uh}.
We find growth at every sampled nonzero spin, with
$\kappa_r=\omega_{\mathrm{ro}}/\Omega$ fitted by
$\kappa_r\simeq0.5238+0.4019\chi^2$ over $0.05\leq\chi\leq0.65$.
Its inertial-frame frequency $|\omega_{\mathrm{in}}|/(2\pi)$ rises to
$1.635$~kHz at $\chi=0.65$, while the growth time shortens from $214$~s at
$\chi=0.3$ to $3.7$~s at $0.65$.}

{Besides the $r$ mode, we follow 18 additional $m=2$ inertial branches,
8 polar-led and 10 axial-led. Figure~\ref{fig:fluid} shows their real
frequencies and the damping or growth times of five representative branches.}
{The ${}^{5}i_1^{\mathrm p}$ branch encounters the prograde $f$ mode near
$\chi=0.529$. A fit of Eq.~\eqref{eq:mode-interaction} gives
$V/(2\pi)\simeq4.58$~Hz and $g/(2\pi)\simeq8.11$~Hz, placing the
encounter in the weak-interaction regime.
The real frequencies cross, while the imaginary parts avoid crossing.
In the $\tau=|\sigma_I|^{-1}$ panel of Fig.~\ref{fig:fluid}, mixing
produces a sharp minimum of $0.112$~s in the inertial-mode damping time
and a local maximum in the $f$-mode damping time.}

Away from mode encounters, we estimate the $f$- and $r$-mode radiation times using
$\Ec\sim\omega_{\mathrm{in}}\mathcal A$, where $\mathcal A$ is the wave action.
Quadrupolar $f$ modes radiate
$P_{\mathrm{GW}}\propto\omega_{\mathrm{in}}^6|Q|^2$, with mass quadrupole $Q$
\cite{Thorne:1980ru}.
Combining the weak-radiation balance
$\sigma_I\simeq P_{\mathrm{GW}}/(2\Ec)$ with the wave action estimated
from an isolated $f$-mode pair gives
$\tau_f^{-1}\simeq k_fMR_e^2|\omega_{\mathrm{in}}|^5/\Delta$, where
$\Delta=|\omega_{\mathrm{in}}-\omega_{\mathrm{in}}'|/2$ is half the separation from the partner
$f$ branch and $R_e$ is the circumferential equatorial radius.
Current-quadrupole emission gives the leading $r$-mode estimate
$\tau_r^{-1}\simeq k_rMR_e^4\Omega^6$
\cite{Lindblom:1998wf}.
The dark gray dash-dotted curves in Fig.~\ref{fig:fluid} evaluate these
estimates using each rotating model's $M$ and $R_e$, with fitted
coefficients $k_f=0.0218$ and $k_r=0.0293$.

\textit{Quasiradial neutral point.---}
The axisymmetric quasiradial $F$ mode determines stability against collapse
and reduces to the fundamental radial oscillation of a nonrotating star
\cite{Chandrasekhar:1964zz,Shapiro:1983du}.
{Along each fixed-$J$ equilibrium
sequence, the turning point is reached at the central rest-mass
density $\rho_{c,\mathrm{TP}}$ where the ADM mass $M(\rho_c)$ is maximum
\cite{Cook:1992flp,Cook:1993qj,Cook:1993qr}.}
This turning point supplies a sufficient
condition for secular instability
\cite{Friedman:1988er,Schiffrin:2013zta}, while the dynamical
collapse threshold requires the physical axisymmetric zero mode
\cite{Takami:2011zc}.
{For $\Gamma=2$ stars at mass shedding, simulations in
Ref.~\cite{Shibata:1999yx} found the sampled models
below the estimated turning-point density dynamically stable and placed
the collapse threshold close to that density.}
We construct the $m=0$ static operator with each fluid element's
specific angular momentum held fixed~\cite{Friedman:2013xza}.
Eliminating the metric and azimuthal velocity perturbations gives a
reduced operator on meridional displacements. Its quasiradial eigenvalue
vanishes at the neutral central density $\rho_{c,\mathrm N}$.

For each fixed-$J$ sequence in Fig.~\ref{fig:neutral}, we independently
compute the axisymmetric quasiradial $F$-mode frequency $\sigma_F$ and
fit $\operatorname{Re}(\sigma_F^2)$ to locate its zero and check the static
neutral point. The upper panel shows the oscillation frequency and growth
rate of this $F$ mode across neutrality for three representative sequences.
The connecting curves interpolate $\operatorname{Re}(\sigma_F^2)$ in central enthalpy
and convert the horizontal coordinate using the EOS. They show the recovered
real frequency on the oscillatory side and imaginary frequency on the
growing side. The sign change of $\operatorname{Re}(\sigma_F^2)$
marks the transition from oscillation to nonoscillatory growth.
In the APR4 example with
$J=4\,M_\odot^2$, the mode still oscillates at $(\sigma_F/2\pi)=94$\,Hz at the
turning point. It reaches zero at
$(\rho_c-\rho_{c,\mathrm{TP}})/\rho_{c,\mathrm{TP}}\simeq8.48\times10^{-4}$
and then becomes a growing mode.
We also include a $\Gamma=2$ polytrope, $p=K\rho^2$ with
$K=100\,M_\odot^2$.
For the EOSs studied here, the neutral points lie
at slightly higher central density than the turning points,
$\rho_{c,\mathrm N}>\rho_{c,\mathrm{TP}}$, at every sampled $J$.
This ordering is in clear tension with Ref.~\cite{Takami:2011zc},
which found the neutral line on the
lower-density side of the turning-point line for a $\Gamma=2$ polytrope.
The relative offset $(\rho_{c,\mathrm N}-\rho_{c,\mathrm{TP}})/\rho_{c,\mathrm{TP}}$
increases monotonically with $J$ for MIT and the polytrope, whereas for APR4 it peaks near
$J=4\,M_\odot^2$ and then decreases.

\begin{figure}[!tb]
 \centering
 \includegraphics[width=\columnwidth]{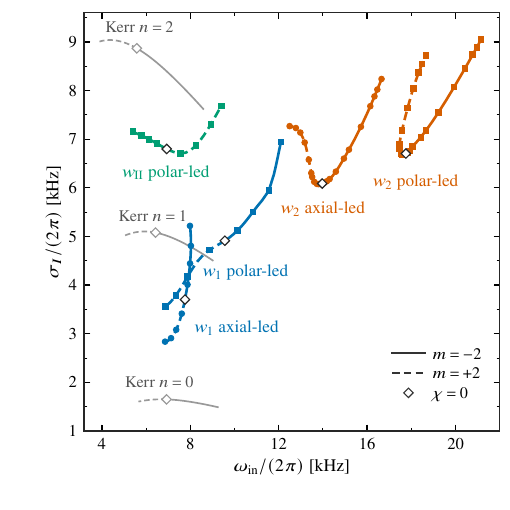}
 \vspace{-10mm}
 \caption{Spacetime-mode trajectories of the APR4 sequence.
Here $w_1$ and $w_2$ denote the fundamental and first radial overtone, respectively.
Solid/dashed curves denote prograde $m=-2$/retrograde $m=+2$ modes for both
stars and Kerr, in the convention of Eq.~\eqref{eq:convention}.
Circles/squares distinguish axial-led/polar-led families. Gray curves show
Kerr $l=2$, $n=0,1,2$ frequencies at the stellar mass and spin.
Diamonds mark nonrotating endpoints.}
 \label{fig:wmodes}
\end{figure}

\textit{Spacetime modes.---}
The same outgoing problem resolves $w$ modes, which are predominantly
spacetime oscillations with little fluid motion
\cite{Kokkotas:1986gd,Kojima:1987nh,Kokkotas:1992lxs}.
Their damping rates are comparable to their real frequencies, testing
the formulation beyond the weak-radiation regime of the fluid modes.
Figure~\ref{fig:wmodes} follows the fundamental $w_1$ and first radial
overtone $w_2$ of the axial-led and polar-led curvature-mode families,
together with the distinct polar-led $w_{\mathrm{II}}$ family
\cite{Leins:1993zz,Andersson:1995ez}.

Over $0\leq\chi\leq0.60$, damping rates increase on all five prograde
branches, with a comparatively small change for $w_{\mathrm{II}}$.
The $w_{\mathrm{II}}$ family exhibits the opposite real-frequency
splitting to $w_1$ and $w_2$ already at low spin: its prograde frequency
decreases while its retrograde frequency increases.
An independent slow-rotation calculation in Regge--Wheeler gauge
reproduces this reversed splitting. At first order, we find
$(\sigma-\sigma_0)/(m\Omega)\simeq-0.84-0.25\ii$ for the polar-led $w_1$
and $+0.70-0.16\ii$ for the polar-led $w_{\mathrm{II}}$, where $\sigma_0$ is
the nonrotating complex frequency.
For both $m=\pm2$ branches at the sampled spins $\chi\leq0.1$,
the real frequencies and damping rates agree with the full calculation
to within $0.7\%$ and $0.3\%$, respectively.

The stellar branches have quality factors
$\mathcal Q=|\omega_{\mathrm{in}}|/(2\sigma_I)<1.4$ in both $m$ sectors
over $0\leq\chi\leq0.60$, with spin dependence that varies among the
families. For all three Kerr modes shown at the same ADM mass and spin,
$\mathcal Q$ increases along the prograde branches and decreases along
the retrograde branches~\cite{Leaver:1985ax}.
The spectra are set by the effective potentials and
boundary conditions~\cite{Andersson:1995ez,Kojima:1992ie}. The exterior
multipole moments enter the wave equations and generally differ between
stars and Kerr black holes of the same mass and spin~\cite{Hartle:1968si}.
A detailed physical explanation of these different rotational
responses is left for future work.

\textit{Summary.---}
Our formulation and numerical implementation enable accurate calculations
of the quasi-normal modes of rapidly rotating relativistic stars in full GR.
Their complex spectra distinguish the CFS
and Jacobi-like neutral conditions from the onset of dynamical bar-mode
growth, while the quasiradial zero mode accurately locates the axisymmetric
collapse threshold. Interactions between $f$ and inertial modes reveal
avoided crossings and exchanges of radiative damping or growth rates.

The next steps are to determine the EOS dependence of these spectra
and instability thresholds and to study the corotation instability
in stars with differential rotation
\cite{Watts:2003nn,Saijo:2005gb,Centrella:2001xp}, which can be straightforwardly done in our code.
However, the theory employed is linear. Following the nonlinear evolution of
CFS-unstable modes in numerical relativity is essential to determine their saturation and
GW emission
\cite{Stergioulas:2000vs,Ou:2004vj,Shibata:2004ub,Lindblom:2000az,Kastaun:2011yd,Yu:2026fsy}.

\textit{Acknowledgments.---}
YG thanks Kostas Kokkotas for discussions on $w$ modes of rotating
stars during his visit to T\"ubingen, which motivated this study of the
complex $w$-mode spectrum.
We used Claude Code and Codex to clarify the Ritz--Galerkin method,
check the derivation of the equations in Lorenz gauge, and help optimize
our numerical codes.
The numerical calculations were performed on the Sakura cluster at
the Max Planck Computing and Data Facility (MPCDF). This work was in part supported by Grant-in-Aid for Scientific Research (grant No.~23H04900) of Japanese MEXT/JSPS.

\bibliography{refs}
\end{document}